\documentclass[10pt,journal]{IEEEtran}
\usepackage{amsmath}
\usepackage{epsfig}
\usepackage{graphicx}
\usepackage{amsmath}
\usepackage{amsfonts}
\usepackage{latexsym}
\usepackage{amssymb}
\usepackage{cite}
\usepackage{array}

\newcommand{\br}{\mathbf{r}}

\newcommand{\by}{\mathbf{y}}

\newcommand{\bv}{\mathbf{v}}
\newcommand{\bq}{\mathbf{q}}

\newcommand{\be}{\mathbf{e}}
\newcommand{\bO}{\mathbf{0}}

\newcommand{\bR}{\mathbf{R}}

\newcommand{\bD}{\mathbf{D}}

\begin{document}

\title{Sparse Digital Cancellation of Receiver Nonlinear Distortion in Carrier Aggregation Systems}
\author{Ahmad Gomaa, Charles Chien, Ming Lei, ChihYuan Lin and Chun-Ying Ma \\
Mediatek Incorporation}

\maketitle

\begin{abstract}
In carrier-aggregation systems, digital baseband cancelation of self-interference generated by receiver nonlinearity requires the estimation of several reference signals contributions. As the nonlinearity order and frequency selectivity of the chip response increase, the number of reference signals significantly increases rendering the estimation of their contributions more complex. We propose a sparsity-based approach for the selection of the reference signals to match the distortion interference using a few reference signals. Simulation results show significant performance improvement over prior art with the same complexity.
\end{abstract}

\setlength{\textfloatsep}{10pt}
\setlength{\floatsep}{8pt}
\section{Introduction}\label{sec_Introduction}
In carrier aggregation (CA) systems, signals are transmitted and/or received over multiple frequency bands simultaneously. CA is adopted in the long-term evolution (LTE) standard \cite{LTE_36101}. Due to imperfect chip and board isolations, one or more uplink (UL) signals can then leak into the low-noise amplifiers (LNAs) of the receive chains of downlink (DL) signals. Practical LNAs exhibit non-linearity behaviour \cite{Razavi-book} generating harmonic distortion (HD) and inter-modulation distortion (IMD) of the leaked UL signals that can lie at the DL band. This problem causes significant self-interference that degrade the performance of the victim band \cite{LTE_36101}.

In \cite{Asbeck_TMTT_2013}, digital baseband interference cancellers are used, where the UL baseband signals were utilized to regenerate the distortion and subtract it from the received signal. Before reaching the LNA input, UL signals are shaped by the chip frequency response. Hence, the LNA-generated distortion is a nonlinear function of the weighted summation of several delayed versions of the original UL signals. The summation weights represent the chip channel response at the corresponding lags. To avoid using nonlinear cancellation filters, the distortion signal is decomposed into the linear combination of the UL signals raised to different powers and delayed by different lags, called reference signals. The reference signals weights are functions of the original chip response and can be estimated using any linear estimation technique \cite{Asbeck_TMTT_2013}. In \cite{Kahrizi_IM2_2008,Omer_MTT_2011} and \cite{Omer_MTT_2012}, digital cancellation of nonlinearity distortion generated at the transmitter and receiver, respectively. In \cite{Omer_MTT_2012}, the IMD was between one modulated signal and an unmodulated tone, unlike this paper and \cite{Asbeck_TMTT_2013} where both inter-modulated signals are modulated. Unlike modulated signals, unmodulated tones do not experience frequency-selective channel.

The reference signals in \cite{Asbeck_TMTT_2013} are static and pre-defined regardless of the chip channel response. Instead, we propose a sparsity-based solution to dynamically select the reference signals, where all candidate reference signals are included in a dictionary matrix. Then, based on their auto-correlation and cross-correlation with the observed signal, a subset of them is selected to represent the distortion signal. The number of selected reference signals is flexibly set based on the design constraints on complexity and power. Unlike \cite{Asbeck_TMTT_2013}, our approach enables different subset selection for different chip responses.
The rest of the paper is organized as follows. The system model and problem formulation are described in Sections \ref{sec_SysModel} and \ref{sec_problem_form_prior_art}, respectively. Our proposed sparsity-based approach is presented in Section \ref{sec_algorithm}. Simulation results are provided in Section \ref{sec_Simulations}, and the paper is concluded in Section \ref{sec_Conclusion}.

\textit{Notations}: Lower and upper case bold letters denote vectors and matrices, respectively, and $\bO$ denotes the all-zero vector. Also, $(\,)^*$, $(\,)^T$ and $(\,)^H$ denote the complex conjugate, transpose and conjugate transpose operations, respectively. The notation $\left|\;\right|$ denotes the absolute value.

\section{System Model}\label{sec_SysModel}
We describe the two main nonlinearity distortion models known in the literature, namely, HD and IMD.

\subsection{Harmonic Distortion Model}\label{sec_HD_model}
In practical transceivers, the UL signal up-converted to carrier frequency $f_\text{tx}$ leaks into the receiver LNA. Due to its inevitable nonlinearity, the LNA generates the $Q$-th order harmonic of the leakage UL signal at frequency $Q\times f_\text{tx}$. In FDD systems where the receiver oscillator frequency $f_\text{rx}=Qf_\text{tx}$, the HD of the UL leakage signal will interfere with the desired downlink (DL) signal de-sensitizing the whole receiver chain. In CA systems, $f_\text{tx}$ and $f_\text{rx}$ are the UL and DL carrier frequencies of two different aggregated bands, for example, bands 12 (710MHz) and 4 (2.13 GHz), respectively, in LTE \cite{LTE_36101}. We denote the baseband time-domain (TD) UL signal at the digital-to-analog converter (DAC) input by $s(n)$. The UL signal leaks to the LNA input through a finite impulse response (FIR) channel representing the chip response $\left\{h(k)\right\}_{k=0}^{L-1}$ where $L$ is the channel length. The TD baseband equivalent of the leakage signal at the LNA input is given by:
\begin{equation}\label{eqn_x_HD}
  x(n) = \sum_{k=0}^{L-1} h(k) s(n-k)
\end{equation}
We write the $Q$-th order HD signal at the analog-to-digital converter (ADC) output
as follows:
\begin{equation}\label{eqn_HD_signal}
    p(n)=c_0 x^Q(n)
\end{equation}
where $c_0$ is related to the $Q^\text{th}$ order input-referred intercept point (IIP) of the LNA \cite{Razavi-book}.


\subsection{Inter-Modulation Distortion Model}\label{sec_IMD_model}
In carrier aggregation, the two UL signals of different frequencies, $f_\text{tx,1}$ and $f_\text{tx,2}$ leak to the receiver LNA. As a result of the LNA nonlinearity, these two leakage signals inter-modulate together creating an IMD signal sitting at a new frequency $pf_\text{tx,1}+qf_\text{tx,2}$, where $p$ and $q$ are integer nonzero numbers. If this new frequency equals to the frequency of the downlink signal $f_\text{rx}$, then the system is said to suffer from IMD. The IMD order is given by $Q_I = |p|+|q|$. In LTE, bands 3 (1750MHz) and 20 (850MHz) can cause $3^\text{rd}$-order IMD ($p=2,q=-1$) to band 7 (2660MHz) \cite{LTE_36101}. We write the $Q_I$-th order IMD signal at the ADC output for $p,q>0$ as follows:
\begin{align}\label{eqn_IMD_p_q_positive}
  p(n) = c_0 x_1^{|p|}(n)  x_2^{|q|}(n)
\end{align}
where $x_1(n)$ and $x_2(n)$ are the TD baseband equivalent of the two UL leakage signals seen at the LNA input, given by:
\begin{align}
  x_1(n) \hspace{-0.1cm}=\hspace{-0.2cm} \sum_{k=0}^{L_1-1}\hspace{-0.1cm} h_1(k) s_1(n-k),\;
  x_2(n) \hspace{-0.1cm}=\hspace{-0.2cm} \sum_{k=0}^{L_2-1}\hspace{-0.1cm} h_2(k) s_2(n-k)\label{eqn_x1_x2_IMD}
\end{align}
where $s_1(n)$ and $s_2(n)$ are the complex baseband TD signals at the inputs of the DACs associated with the two UL chains. Furthermore, $h_1(k)$ and $h_2(k)$ are the FIR channels of lengths $L_1$ and $L_2$, respectively, representing the chip responses between the two UL chains and the LNA input. The IMD expression in \eqref{eqn_IMD_p_q_positive} can be also written for $p<0$ and/or $q<0$ with the following modification. If $p$ or $q$ is negative, we replace the corresponding signal $x_1(n)$ or $x_2(n)$, respectively, in Equation \eqref{eqn_IMD_p_q_positive} by its complex conjugate.

\section{Problem Formulation and prior Art}\label{sec_problem_form_prior_art}
\subsection{Problem Formulation}\label{sec_problem_formulation}
The distortion signal $p(n)$ in \eqref{eqn_HD_signal} or \eqref{eqn_IMD_p_q_positive} for HD or IMD schemes, respectively, interferes with the desired DL signal $y(n)$ of power $P_s$ yielding:
\begin{equation}\label{eqn_rx_signal}
  r(n) = p(n) + y(n) + z(n)
\end{equation}
where $z(n)$ is the complex background additive Gaussian (AWG) noise of single-sided power spectral density $N_o$. The cancellation algorithm exploits the knowledge of the transmitted UL signal $s(n)$ (or $s_1(n)$ and $s_2(n)$ in IMD cases) to constructs the distortion signal $p(n)$ and cancel it from $r(n)$. The algorithm requires the estimation of the FIR channel $h(k)$ (or $h_1(k)$ and $h_2(k)$) representing the chip response to construct the leakage signal $x(n)$ (or $x_1(n)$ and $x_2(n)$) seen at the LNA input. However, the observed signal $r(n)$ is not linear in the unknown channel $h(k)$ due to the LNA nonlinearity in \eqref{eqn_HD_signal} and \eqref{eqn_IMD_p_q_positive}. Hence, we need to expand the polynomials of the distortion signal, e.g., $\left(\sum_{k=0}^{L-1} h(k) s(n-k)\right)^Q$ and $\left(\sum_{k=0}^{L_1-1} h_1(k) s_1(n-k)\right)^{|p|}\left(\sum_{k=0}^{L_2-1} h_2(k) s_2(n-k)\right)^{|q|}$, using the following multinomial theorem:
\begin{equation}\label{eqn_multinomial}
  \left(\sum_{k=1}^{K} a_k\right)^N = \sum_{\substack{t_1+t_2+\ldots+t_N = N,\\t_1,t_2,..,t_N=0,1,2,..}}^{} \frac{N!}{t_1!t_2!\ldots t_N!} \prod_{1\leq k  \leq N}^{} a_k^{t_k}
\end{equation}
For instance, the first polynomial in the right-hand side (RHS) of \eqref{eqn_HD_signal} is expanded for $Q=3$ and $L=2$ as follows:
\begin{align}\label{eqn_expansion_example}
  c_0 x^Q(n)\hspace{-0.1cm}=\hspace{-0.1cm} c_0\hspace{-0.15cm} \left(\sum_{k=0}^{1} h(k) s(n\hspace{-0.1cm}-\hspace{-0.1cm}k)\hspace{-0.1cm}\right)^3 \hspace{-0.2cm}=\hspace{-0.15cm} \sum_{t=0}^{3} \hspace{-0.07cm}\bar{h}_t s^t(n)s^{3-t}(n\hspace{-0.1cm}-\hspace{-0.1cm}1)
\end{align}
where $\left\{\bar{h}_t\right\}_{t=0}^{3}$ are the new parameters to be estimated. Rewriting Eqn. \eqref{eqn_rx_signal} in matrix-vector format, we get
\begin{equation}\label{eqn_system_equation}
  \br = \bD \bv + \be
\end{equation}
where $\br=\left[r(0),r(1),\ldots,r(P-1)\right]^T$, and $P$ is the number of observed samples used in the estimation process. Furthermore, the columns of the matrix $\bD$ represent the distortion reference signals obtained by the expansion of the distortion polynomials. Following the example in \eqref{eqn_expansion_example}, the $t$-th columns of $\bD$ is filled by the samples $\left\{s^t(n)s^{3-t}(n-1)\right\}_{n=0}^{P-1}$. The length-$\bar{L}$ vector $\bv$ contains the new parameters to be estimated representing the contributions of the reference signals in the columns of $\bD$. Finally, the vector $\be$ is the error vector containing the desired DL signal $y(n)$, background noise $z(n)$, and the part of the distortion signal not represented in the columns of $\bD$ due to its weak contribution. The linear least squares (LLS) solution of $\bv$ is given by the solution of the following minimization problem \cite{Kay}:
\begin{equation}\label{eqn_normal_equations}
  \hat{\bv} = \underset{\bar{\bv}}{\text{min}} \left\|\br-\bD\bar{\bv}\right\|^2  =  \bR^{-1} \bq
\end{equation}
where $\bR = \bD^H\bD$ is the auto-correlation matrix of the reference signals. Moreover, $\bq = \bD^H\br$ represents the cross-correlation vector between the observed signal and the reference signals. Transforming the nonlinear estimation problem into a linear one comes at the expense of increasing the problem dimension, especially for high distortion levels ($Q$ and $Q_I$) and channel length $L$. In HD schemes, modeling only $x^Q(n)$ in the reference matrix $\bD$ increases the problem dimension from $L$ to $\bar{L} = \frac{(Q+L-1)!}{Q!(L-1)!}$. For $Q=3$ and $L=4$, this corresponds to increasing the problem dimension from $L=4$ to $\bar{L} = 20$. The new problem dimension (after multinomial expansion) becomes even larger for IMD schemes.
In Section \ref{sec_algorithm}, we show how sparsity-based techniques are used to reduce the number of parameters to be estimated.

\subsection{Prior Art and other approaches}\label{sec_prior_art}
The prior-art IMD cancellation approach in \cite{Asbeck_TMTT_2013} models the $3^\text{rd}$-order IMD reference signal as follows:
\begin{equation}\label{eqn_Asbeck_IMD_model}
  p_\text{prior}(n)\hspace{-0.1cm} = \hspace{-0.15cm}\left(\sum_{k_1=0}^{L_1-1}\hspace{-0.15cm} h_1(k) s_1^*(n\hspace{-0.1cm}-\hspace{-0.1cm}k_1)\hspace{-0.1cm}\right)\hspace{-0.2cm} \left( \sum_{k_2=0}^{L_2-1}\hspace{-0.15cm} h_2(k_2) s_2^2(n\hspace{-0.1cm}-\hspace{-0.1cm}k_2) \hspace{-0.1cm}\right)
\end{equation}
where the total number of terms in the expansion of \eqref{eqn_Asbeck_IMD_model} is $J_\text{prior}=L_1\times L_2$. Comparing \eqref{eqn_Asbeck_IMD_model} with the exact representation in Eqns. \eqref{eqn_IMD_p_q_positive} and \eqref{eqn_x1_x2_IMD}, we find the model in \cite{Asbeck_TMTT_2013} is different from the exact model. The model in \cite{Asbeck_TMTT_2013} squares $s_2(n-k_2)$, while the squaring should go over $s_2$ \textit{after} passing through the filter $h_2(k)$, i.e., the squaring should go over $\left( \sum_{k_2=0}^{L_2-1} h_2(k_2) s_2(n-k_2) \right)$. The IMD reference signal was modeled in \cite{Asbeck_TMTT_2013} \textit{as if} $s_2(n)$ go through the non-linearity \textit{before} getting filtered which is not true as discussed in Section \ref{sec_SysModel}.

Another candidate approach would be to model the distortion \textit{as if} the nonlinearity is generated \textit{before} the channel. We call it the Hammerstein-based approach as it follows the well-known Hammerstein model \cite{Hammerstein}. For example, the first nonlinearity term in the RHS of \eqref{eqn_IMD_p_q_positive} modeling the IMD signal is approximated by:
\begin{equation}
  p_\text{Hammerstein}(n) = \sum_{k=0}^{J_\text{Hamm}-1} h(k) s_1^{|p|}(n-k)  s_2^{|q|}(n-k)
\end{equation}
For both the prior-art and Hammerstein-based approaches, the corresponding reference matrix ($\bD_\text{prior}$ or $\bD_\text{Hamm}$) will have $J_\text{prior}$ or $J_\text{Hamm}$ columns, respectively. Then, the LLS-based estimation of the distortion signature is then obtained for the prior-art approach (and similarly for the Hammerstein-based approach) as follows:
\begin{equation}\label{eqn_LLS_prior_art}
  \hat{\bv}_\text{prior} = \left(\bD_\text{prior}^H\bD_\text{prior}\right)^{-1}\bD_\text{prior}^H\br
\end{equation}


\section{Proposed Algorithm}\label{sec_algorithm}
We propose a novel approach for improved reference signal design while controlling the problem dimension. Our approach comprises two main steps. First, we use the multinomial expansion in \eqref{eqn_multinomial} and include all the reference signals out of this expansion in the reference matrix $\bD$. The second step is to obtain a $J_\text{sparsity}$-sparse solution of $\bv$ in \eqref{eqn_system_equation} with only $J_\text{s}<<\bar{L}$ nonzero entries. This sparse solution $\hat{\bv}_s$ should keep $\bD\hat{\bv}_s$ close to the observation vector $\br$. This requirement can be casted into the following optimization problem:
\begin{equation}\label{eqn_L0_problem}
  \hat{\bv}_s = \underset{\bar{\bv}}{\text{min}} \left\|\br-\bD\bar{\bv}\right\|^2 \quad\text{subject to}\quad \left\| \bar{\bv} \right\|_0 = J_\text{s}
\end{equation}
where $\left\| \bar{\bv} \right\|_0$ is the $l_0$-norm of the argument vector and represents the number of nonzero entries in this vector. However, the solution of \eqref{eqn_L0_problem} requires an intensive computation burden even for moderate values of $\bar{L}$ and $J_\text{s}$ due to the huge search space. Several techniques have been proposed in the literature to obtain approximate solutions of \eqref{eqn_L0_problem}, e.g., Orthogonal Matching Pursuit (OMP) \cite{OMP}, Orthogonal Least Squares (OLS) \cite{OLS}, and FOCUSS \cite{FOCUSS}.
We choose the OMP thanks to its implementation simplicity and efficiency. The OMP algorithm takes as input both the reference matrix $\bD$, the observation vector $\br$, and the required sparsity level $J_\text{s}$. The OMP output is an approximate $J_\text{s}$-sparse solution $\hat{\bv}_\text{OMP}$ of \eqref{eqn_normal_equations} as follows:
\begin{equation}\label{eqn_OMP_input_ouput}
  \hat{\bv}_\text{OMP} = \text{OMP}\left(\bD,\by,J_\text{s}\right)
\end{equation}
The OMP technique is described as follows:

{\bf Initialization}: Define an empty index set $I_0=\phi$, set the initial residual $\br_0=\by$, initialize $\hat{\bv}_\text{OMP}=\bO$, and set $k=1$.

{\bf The $k^\text{th}$ iteration}:
\begin{enumerate}
\item \label{step1} Compute $\delta_i=\left|\br_{k-1}^H\bD(:,i)\right|$ for all $i \notin I_{k-1}$.
\item \label{step2} Choose $c_k=\underset{i}{\operatorname{arg}}\;\underset{i}{\operatorname{max}}\;\delta_i$.
\item \label{step3} Update $I_k=I_{k-1} \cup c_k$. In this step, the indices of the nonzero elements are augmented by $c_k$, the index of the $k^\text{th}$ nonzero entry computed at the $k^\text{th}$ iteration.
\item \label{step4} Compute \begin{equation}\label{eqn_OMP1}
                              \hat{\bv}_\text{OMP}(I_k)\hspace{-0.1cm}=\hspace{-0.1cm}\left(\hspace{-0.1cm}\left(\bD(:,I_k)\right)^H\bD(:,I_k)\hspace{-0.1cm}\right)^{-1}\hspace{-0.1cm}\left(\bD(:,I_k)\right)^H \hspace{-0.1cm}\by
                            \end{equation}
                            where $\hat{\bv}_\text{OMP}(I_k)$ holds $\hat{\bv}_\text{OMP}$ elements indexed by $I_k$.
\item \label{step5} Compute $\br_k=\by-\bD(:,I_k)\hat{\bv}_\text{OMP}(I_k)$ where $\br_k$ is the residual error term at the $k^\text{th}$ iteration.
\item \label{step6} If $k = J_\text{s}$, exit the algorithm, else set $k=k+1$ and go to Step \ref{step1}.
\end{enumerate}
In words, OMP tries to find the columns (atoms) of the matrix $\bD$ (dictionary) whose linear combination is close (matched) to $\by$. From \eqref{eqn_OMP1}, we find that the maximum size of the matrix to be inverted is $J_\text{s}$, controlled by the designer and can be made much smaller than $\bar{L}$. We can actually set $J_\text{s}=L$ and let the OMP technique choose the $L$ reference vectors that closely matches the observed signal. The selected set of reference vectors are adaptive and can be different from a channel response to another. This is clearly different from \cite{Asbeck_TMTT_2013} where the reference signals are pre-set regardless of the channel response. The matrix inversion in \eqref{eqn_OMP1} is efficiently implemented using the Cholesky decomposition algorithm \cite{Golubbook}. An efficient OMP implementation algorithm using adaptive Cholesky decomposition is proposed in \cite{OMP_Cholesky_89,OMP_Cholsky_99}, where the decomposition step is not performed every iteration. Instead, it is observed that the matrix $\left(\bD(:,I_k)\right)^H\bD(:,I_k)$ is the same as the matrix $\left(\bD(:,I_{k-1})\right)^H\bD(:,I_{k-1})$ in the previous iteration except for an extra augmented row and column. This observation saves computations by using the same Cholesky decomposition of the previous iteration with one more augmented vector obtained by forward substitution \cite{OMP_Cholesky_08}.

\section{Simulation Results and Discussion}\label{sec_Simulations}
\begin{figure}
\centering
\epsfig{file=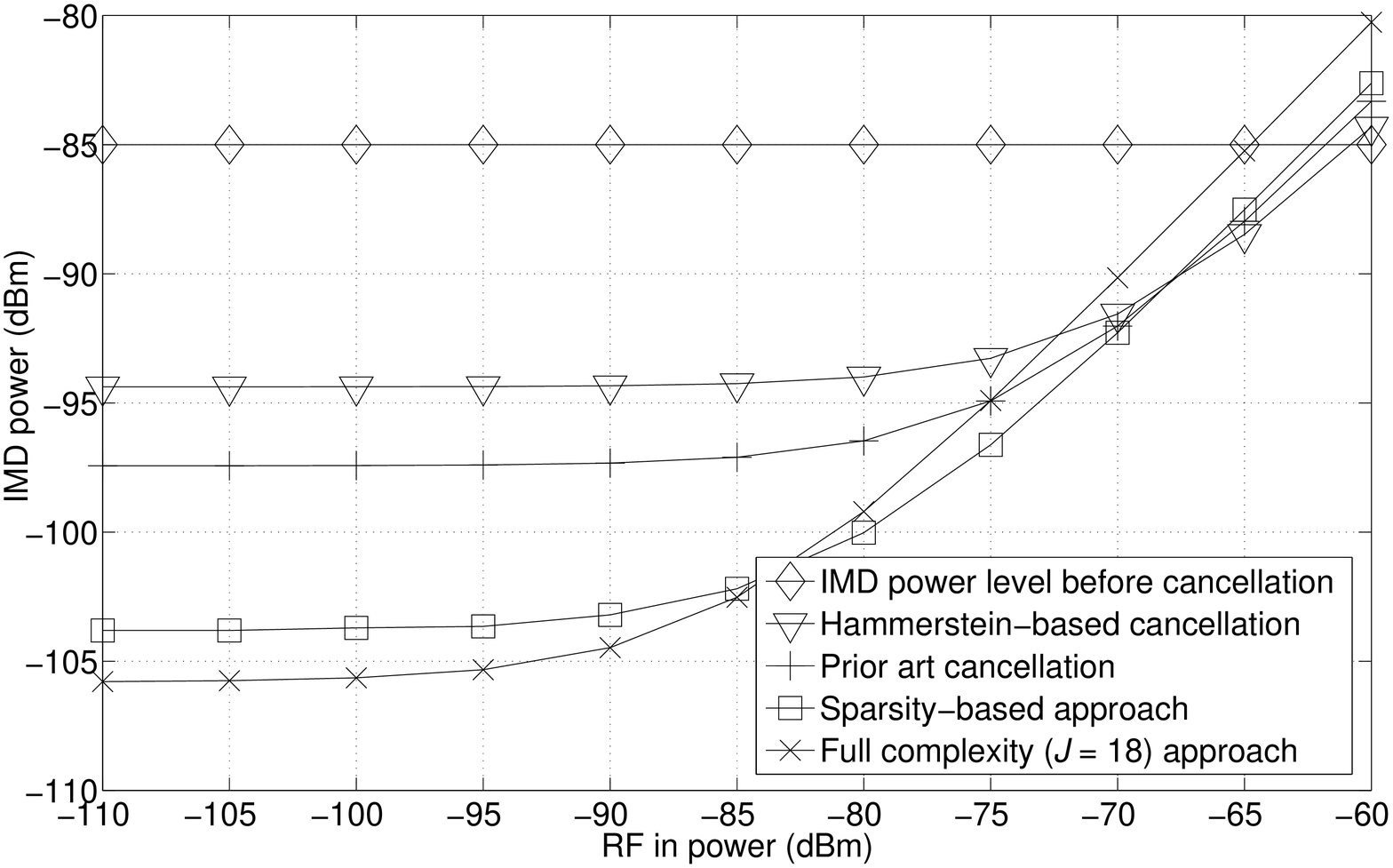,height=2.0in,width=3.4in}
\caption{Original and residual IMD power levels for different cancellation approaches versus DL signal power.} \label{fig_IMDpwr_Ps}
\end{figure}

We simulate the performance of our proposed approach and compare it with the other approaches described in Section \ref{sec_prior_art}. We use practical chip responses provided by well-known manufacturers. In Fig. \ref{fig_IMDpwr_Ps}, we simulate the cancellation performance for the $3^\text{rd}$-order IMD mechanism generated by bands 3 and 20 on band 7 as described in Section \ref{sec_IMD_model}. The uplink signals of bands 3 and 20 both have bandwidths of 10 MHz, and the receive bandwidth of band 7 is also 10 MHz. The block size $P$ is set to 520 samples. The IMD level before cancellation is -85 dBm while the IMD-to-noise power ratio  (INR) is set to 0 dB. Fig. \ref{fig_IMDpwr_Ps} shows the IMD power levels before and after cancellation over a wide range of the DL receive signal power $P_s$. Using the same number of cancellation taps $J_\text{s}=J_\text{prior}=J_\text{Hamm}=9$, our sparsity-based approach shows 6 and 9 dB improvements over the prior-art approach in \cite{Asbeck_TMTT_2013} and the Hammerstein-based approach, respectively, described in Section \ref{sec_prior_art}. For our sparsity-based approach, we construct the columns of the dictionary matrix $\bD$ by setting $L_1=L_2=3$ in \eqref{eqn_IMD_p_q_positive} and \eqref{eqn_x1_x2_IMD}, hence, $\bD$ has 18 columns. For the prior-art approach in \cite{Asbeck_TMTT_2013}, we set $L_1=L_2=3$, so $J_\text{prior}=L_1\times L_2=9$. In Fig. \ref{fig_IMDpwr_Ps}, we also show the performance of our canceller with full complexity, i.e., $J = \bar{L} = 18$, where all columns of $\bD$ are utilized in IMD estimation and cancellation. Our approach shows additional 6 dB interference suppression over prior art in \cite{Asbeck_TMTT_2013}. As expected, the residual IMD power level gets higher than that of the original IMD at high DL signal power $P_s$. The reason is the poor estimation accuracy due to low IMD to DL signal power ratio (ISR). However, in practice the original IMD power will not be fixed to -85 dBm regardless of $P_s$ as in Fig. \ref{fig_IMDpwr_Ps}. Instead, it decreases as $P_s$ increases because the user equipment (UE) gets closer to the base station, and the UL signal is transmitted at lower power levels.

\begin{figure}
\centering
\epsfig{file=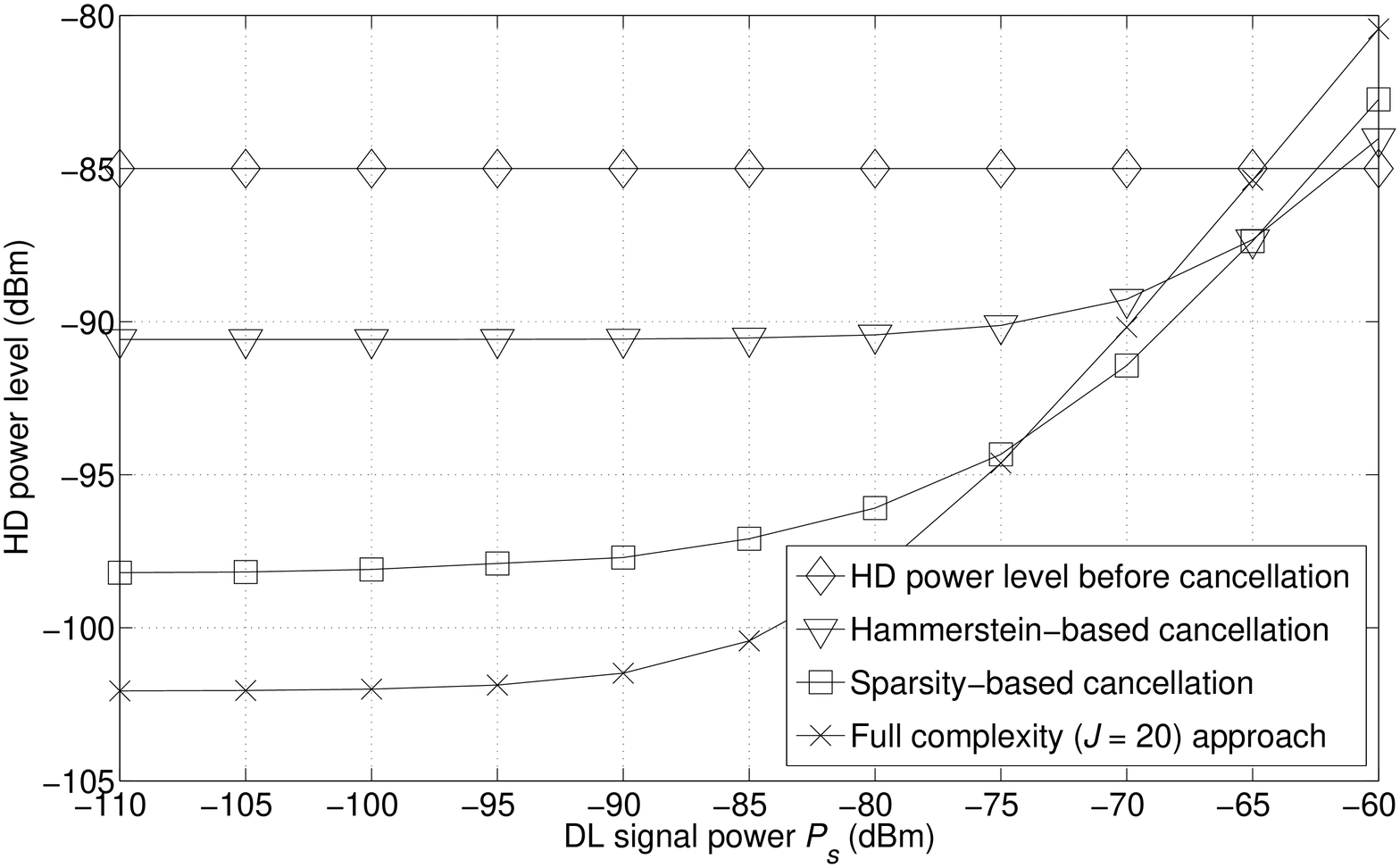,height=2.0in,width=3.4in}
\caption{Original and residual HD power levels for different cancellation approaches versus DL signal power.} \label{fig_HDpwr_Ps}
\end{figure}

\begin{figure}
\centering
\epsfig{file=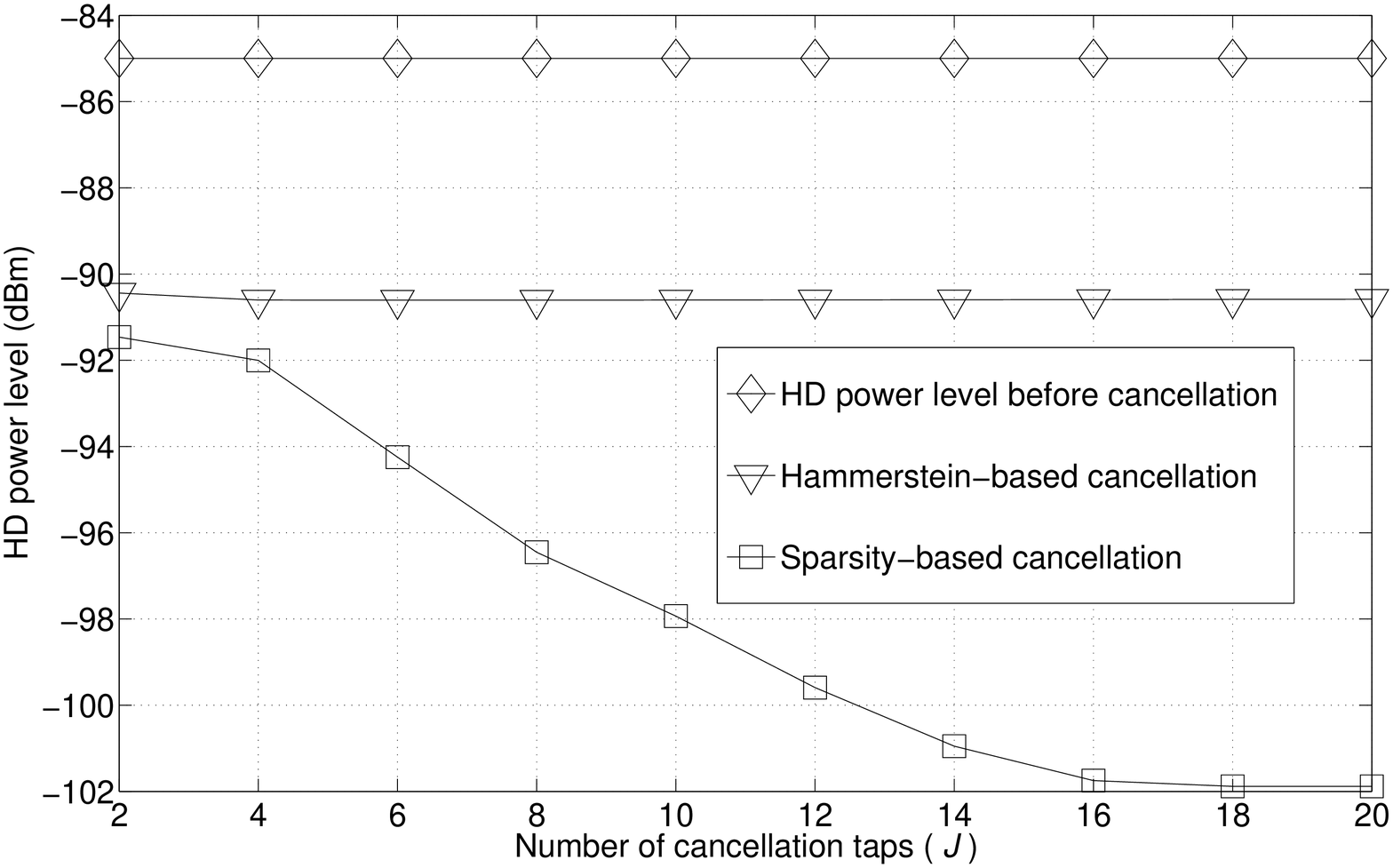,height=2.0in,width=3.4in}
\caption{Original and residual HD power levels for different cancellation approaches versus number of cancellation taps.} \label{fig_HDpwr_numtaps}
\end{figure}

In Figs. \ref{fig_HDpwr_Ps} and \ref{fig_HDpwr_numtaps}, we simulate the performance for the $3^\text{rd}$-order HD mechanism in Section \ref{sec_HD_model} where band 12 generates interference on band 4. The UL and DL bandwidths of bands 12 and 4, respectively, are both set to 10MHz. The original HD power level is set to -85 dBm with INR $=0$ dB and $P=520$ samples. Since the prior art in \cite{Asbeck_TMTT_2013} was proposed for IMD but not for HD, we compare our sparsity-based approach with the Hammerstein-based approach. For our sparsity-based approach, the columns of the dictionary matrix $\bD$ is constructed by setting $L=4$ in \eqref{eqn_x_HD} and \eqref{eqn_HD_signal}, hence $\bD$ has 20 columns. In Fig. \ref{fig_HDpwr_Ps}, the residual HD power level is simulated for both approaches using $J=10$ cancellation taps over a wide range of $P_s$. The performance of the full complexity ($J = 20$) is also shown. In Fig. \ref{fig_HDpwr_numtaps}, we fix $P_s = -95$ dBm and compare the performances of both approaches for different number of cancellation taps $J$. The superiority of our sparsity-based approach is clear, where increasing $J$ improves its performance as it improves the HD estimation through including more component reference vectors in the estimation process. However, increasing $J$ does not improve the performance of the Hammerstein-based approach because its estimation process does not include the contribution by most of the cross terms of the HD expansion equation, c.f. Eqn. \eqref{eqn_expansion_example}.

\section{Conclusion}\label{sec_Conclusion}
We proposed a sparse linear filter to cancel the distortion signal generated by the HD and IMD of uplink signals leaking into the receiver LNA. The reference signals selection process is dynamic and adapts itself for different channel responses of the chip through correlating the observed vector with the dictionary reference signals. Our sparsity-based approach provides additional interference suppression over the prior-art approach using the same number of filter taps.

\bibliographystyle{IEEEtran}
\bibliography{IEEEabrv,reference}
\end{document}